# A SIMPLE HYPOTHESIS ON THE ORIGIN AND PHYSICAL NATURE OF QUANTUM SUPERPOSITION OF STATES

Pavel V. Kurakin[1], George G. Malinetskii[1]

[1] Keldysh Institute of Applied Mathematics, Russian Academy of Sciences, Moscow
mailto: kurakin.pavel@gmail.com

## 1

We start from considering free electromagnetic field. We suggest noting that propagation and scattering (still not emitting) of electromagnetic field in quantum domain brings no new modes of motion compared to classical domain, in the following sense.

Let us assume standard double-slit experiment for monochromatic light ("the heart of quantum mechanics", by R. Feynman). Probabilities pattern just repeats field intensity of classical light.

So we mean that equations of classical optics actually provide full phase space for single light quanta. The correspondence is mutually single-valued: any screen point reachable by classical light is reachable for single quanta, and vice versa.

Applying this observation to equations for charges can not be done, at least, in direct way. Quantized levels of hydrogen-like atom have no direct analogue in classical domain.

On the other side, an atom does not consist of charges only; it's a compound system of charges and quantized electromagnetic field. This complicates the situation, but we suggest not thinking of this.

We simply suggest considering only light because it is the only thing that makes charges to interact.

## 2

So, actually we have almost full equivalence of quantum and classical propagation of light, excepting the fact that quantum light is quantum, i.e. only fixed (for definite wavelength) portion of energy can be absorbed.

Here we come to our hypothesis itself. Classical Maxwell equations *for free electromagnetic field* can possibly mean much more then they are currently accepted to mean, within classical domain only. These equations can quite be fully applicable in quantum domain as well, but the *physical sense* of these equation changes.

Propagation of light as a wave in accordance to these equations (thus *providing superposition of states*) means some kind of self-organizational process. This process makes some investigation of all possibilities in space-time, **but** *it does not take place in classical space-time*. In other words, the physical time does not "tick" while this process evolves.

Waves performing self-organized search can be referred to as "scout waves" to distinguish them from classical waves which obey *the same* equations.

The process eventually finishes at some detector (screen point) with probability proportional to classical light intensity, which fully corresponds to quantum mechanics. This point event means *emergence* of physical (classical) time instant.

## 3

Note that it is a very general scheme. In [1] we suggested *one possible way* to construct such process. We did not mention Maxwell equations properties there but we started from Feynman's many-paths formulations of quantum theory.

Note that exploring of all possibilities for emitted light quantum implies some kind of back propagation. Both Maxwell's equations and Feynman's path integrals admit back-in-time propagation.

The first possibility is used by John Cramer in transactional interpretation of quantum mechanics [2], while the second is used in [3] to simplify calculating Feynman path integrals.

## 4

In [1] we introduced the notion of "hidden time". Self-organization of light quanta, resulting in falling onto some certain detector, evolves in this hidden time. To stay consistent with usual physical (classical) time, we also introduce a "sewing procedure", which defines what instants of hidden time correspond to physical time. We show that classical time stays a *fully relativistic* notion after this reconsideration.

We realize that such a sewing procedure is somewhat artificial. It looks as if a charge that scatters light is described by some additional discrete variable like a flag. This flag is "on" when the physical time ticks at its locus and is "off" otherwise.

But we believe that actually there can be no need in such a flag variable. It is known that microscopic events are actually detected by some amplifier. Say, an avalanche of electrons emerges in photo – electronic multiplier.

The theory can be built in such a way that detected events correspond to a kind of avalanches, i.e. some collective and self - amplifying modes for charges, starting from single electron scattering scout wave. In this case we have no need to distinguish hidden time and physical time. Physical (macroscopical) time emerges here from microscopical time.

When emerged, physical time loses back propagation, but in principle there's no any fundamental difference to microscopic time.

## 5

As mathematicians we believe that standard quantum theory needs a foundation in terms of dynamical systems approach, because dynamical systems is the mainstream of **all** science and QM should be returned in the "pale of the church".

Saying that this is "interpretational problem" seems senseless to us, because only time can tell which is interpretation only and which is a *new theory*.

We believe that we suggest the *simplest* step in this direction. It is very important not to omit the simplest models even if they are mistaken. The $1^{st}$ nonlinear model of turbulence by L.D. Landau [4] was extremely simple and it turned out to be wrong. And this fallacy was of great significance! Science **must** be sure it does not miss simplest ways in building models.

This is why we invite all theoretical physicists to show us where we are deliberately wrong.

**References**

1. P.V. Kurakin, G.G. Malinetskii, H. Bloom. "*Conversational (dialogue) model of quantum transitions*", quant-ph/0504088.
2. John G, Cramer. "*Transactional interpretation of quantum mechanics*". Reviews of Modern Physics, No. 58, pp. 647 – 688, 1986.
3. G.N. Ord. "*The Feynman Propagator from a Single Path*". Physical Review Letters, Vol. 89, No. 25, 16 December 2002.
4. L.D. Landau. "*To the problem of turbulence*". USSR Academy of Sciences Reports, No. 44, pp. 30-39, 1944.